\pdfoutput=1
\documentclass[sigconf]{acmart}
\usepackage{multirow}
\usepackage{footnote}
\usepackage{graphicx}
\usepackage{booktabs,caption}
\usepackage[flushleft]{threeparttable}
\usepackage{algorithm}
\usepackage{algorithmicx}
\usepackage{algpseudocode}
\usepackage{amsmath}
\usepackage{algpseudocode}
\usepackage{mathrsfs}
\usepackage{pifont}
\usepackage{url}
\usepackage{makecell}
\usepackage{subcaption}
\usepackage{enumitem}
\usepackage{balance}
\usepackage[usenames,dvipsnames]{colortbl}\usepackage{array}

\makeatletter
\newcommand{\thickhline}{%
    \noalign {\ifnum 0=`}\fi \hrule height 1pt
    \futurelet \reserved@a \@xhline
}
\definecolor{mycyan}{cmyk}{.8, 0, 0, 0}
\definecolor{gray-0}{rgb}{ .94, 1,1}
\definecolor{gray-1}{rgb}{ .90, 1,1}
\definecolor{gray-2}{rgb}{ .84, 1,1}

\acmYear{2025}
\copyrightyear{2025}
\graphicspath{{figures/}{figures/performance/}}
\begin{document}

\title{PIRA: Pan-CDN Intra-video Resource Adaptation\\ for Short Video Streaming}
\settopmatter{authorsperrow=4}
\author{Chunyu Qiao}
\authornote{These authors contributed equally to this work.}
\affiliation{%
  \institution{Douyin, ByteDance Inc.}
  \city{Beijing}
  \country{China}}
\email{qiaochunyu@bytedance.com}

\author{Tong Liu}
\authornotemark[1]
\affiliation{%
  \institution{Douyin, ByteDance Inc.}
  \city{Beijing}
  \country{China}}
\email{tongliu@bytedance.com}

\author{Yucheng Zhang}
\affiliation{
  \institution{Computer Science, Beijing University of Posts and Telecommunications}
  \city{Beijing}
  \country{China}
}
\authornote{Zhang's work is done during intership at ByteDance.}
\email{zhangyucheng@bupt.edu.cn}

\author{Zhiwei Fan}
\affiliation{%
  \institution{Douyin, ByteDance Inc.}
  \city{Beijing}
  \country{China}}
\email{fanzhiwei.rice@bytedance.com}

\author{Pengjin Xie}
\authornote{Corresponding author. Email: xiepengjin@bupt.edu.cn}
\affiliation{%
  \institution{Artificial Intelligent School, Beijing University of Posts and Telecommunications}
  \city{Beijing}
  \country{China}}
\email{xiepengjin@bupt.edu.cn}

\author{Zhen Wang}
\affiliation{%
  \institution{Douyin, ByteDance Inc.}
  \city{Beijing}
  \country{China}}
\email{wangzhen3560@bytedance.com}

\author{Liang Liu}
\affiliation{
  \institution{Artificial Intelligent School, Beijing University of Posts and Telecommunications}
  \city{Beijing}
  \country{China}}
\email{liangliu@bupt.edu.cn}

\renewcommand{\shortauthors}{Chunyu Qiao et al.}
\begin{abstract}
In large-scale short-video platforms, CDN resource selection plays a critical role in maintaining users' Quality of Experience (QoE) while controlling escalating traffic costs.
To better understand this phenomenon, we conduct in-the-wild network measurements during video playback in a production short-video system.
The results reveal that CDNs delivering higher average QoE often come at greater financial cost, yet their connection quality fluctuates even within a single video—underscoring a fundamental and dynamic trade-off between QoE and cost.
However, the problem of sustaining high QoE under cost constraints remains insufficiently investigated in the context of CDN selection for short-video streaming.
To address this, we propose PIRA, a dynamic resource selection algorithm that optimizes QoE and cost in real-time during video playback.
PIRA formally integrating QoE and cost by a mathematical model, and introduce a intra-video control-theoretic CDN resource selection approach which can  balance QoE and cost under network dynamics.
To reduce the computation overheads, PIRA employs state-space pruning and adaptive parameter adjustment to efficiently solve the high-dimensional optimization problem.
In large-scale production experiments involving $450,000$ users over two weeks, PIRA outperforms the production baseline, achieving a $2.1\%$ reduction in start-up delay, $15.2\%$ shorter rebuffering time, and $10\%$ lower average unit traffic cost, demonstrating its effectiveness in balancing user experience and financial cost at scale.
\end{abstract}

\maketitle

\section{Introduction}\label{introduction}
The explosive growth of short-video streaming platforms has posed significant challenges in delivering high-quality streaming experiences at scale.
Content delivery networks (CDNs), as the critical infrastructure for video distribution, play a pivotal role in ensuring low-latency content delivery to billions of users.
In recent years, video platforms have integrated cost-efficient alternatives to traditional CDNs by leveraging heterogeneous edge resources, such as edge computing nodes, IoT devices, and repurposed servers, into their distribution networks.
This paradigm, embraced by leading short-video platforms like Instagram Reels, has spurred research on peer-assisted content delivery (PCDN) systems~\cite{ye2024kepc,zhang2024enhancing,ma2024sdpush}.
We refer to all types of network resources as pan-CDN (panoramic CDN).
These pan-CDN types, operated by distinct vendors with pricing determined via negotiated commercial agreements, exhibit cost variations primarily driven by resource type rather than vendor-specific discounts.
To our knowledge, existing studies lack a systematic analysis of how diverse pan-CDN types impact QoE and traffic costs in short video streaming.

Based on the platform's pricing tiers, pan-CDNs can be categorized into different classes.
In this work, for example, we use four categories:  \textbf{pan-CDN1} (highest cost), \textbf{pan-CDN2}, \textbf{pan-CDN3}, and \textbf{pan-CDN4} (lowest cost).
Generally, higher-priced pan-CDNs tend to offer better average service quality.
However, optimal pan-CDN selection that relies on average service quality, rather than temporal service quality influenced by real-world network dynamics, results in a performance gap.

To address this gap, we collect one week of video viewing traces from $1$ million users in our short video platform, i.e., Douyin, characterizing session-level network throughput dynamics.
Our findings reveal a time-varying correlation between pan-CDN pricing tiers and service quality.
The most expensive pan-CDN delivers optimal throughput for only $65\%$ of users, highlighting user-specific performance divergence.
Moreover, the network dynamics of pan-CDNs unfold on the timescale of tens of seconds, reflecting rapid fluctuations in throughput.

Prior academic studies~\cite{jiang2016cfa,meng2022dig} primarily focus on improving user QoE leveraging server-side pan-CDN selection, as summarized in Table~\ref{tab:methodcompare},
These approaches select a single "optimal" pan-CDN for each video at its playback session initialization based on the historical bandwidth of available pan-CDNs, aiming to optimize QoE metrics like rebuffering rate and start-up delay.
While effective, they overlook the inherent financial cost differences across pan-CDN providers.
This distinction is critical for large-scale platforms, where traffic expenses vary significantly between premium commercial CDNs and low-cost edge nodes.
Industry solutions aim to select the cheapest pan-CDN resources possible while maintaining user QoE.
However, these approaches typically treat QoE and cost as independent objectives, lacking a systematic framework for formal QoE-cost tradeoff analysis.
More importantly, none of the existing approaches considers intra-session network dynamics.
During a video downloading session, the optimal pan-CDN bandwidth may fluctuate.
In particular, evolving preload algorithms—such as those that prefetch subsequent video chunks to support seamless user swiping interactions—can extend session duration, thereby rendering initial pan-CDN choices suboptimal.
This consequently highlights the necessity of intra-session pan-CDN adaptation in short video streaming.

In this paper, we propose PIRA, the first study to address intra-video session pan-CDN resource selection, aiming to optimize user QoE while reducing delivery costs.
We formally construct a mathematical model tailored to short video playback, enabling video segment-level modeling of QoE and bandwidth costs across different pan-CDN types.
The model additionally captures the potential throughput degradation caused by frequent pan-CDN switching.
Based on this formulation, PIRA adopts a Model Predictive Control (MPC) framework~\cite{mpc} to jointly determine both the pan-CDN type and the duration of each video segment to download (e.g., 2s or 4s) in real time.
By planning several segments ahead and leveraging throughput predictions, PIRA optimizes pan-CDN selection and adaptively balances user QoE and delivery cost under volatile network conditions.


However, the substantial state space introduces significant computational overhead, making the problem non-trivial to solve in real time.
To mitigate this, we prune PIRA’s search space through two domain-specific strategies: pan-CDN resource filtering and download duration pruning.
Specifically, PIRA does not exhaustively explore all possible pan-CDN options and segment durations. Instead, based on the estimated throughput of each pan-CDN, PIRA filters out suboptimal pan-CDN candidates and avoids exploring unnecessarily short download durations.
These pruning techniques significantly improve runtime efficiency without sacrificing decision quality.


\begin{table}[t]
\begin{center}
 \caption{Existing approaches for pan-CDN selection}
\begin{tabular}{ c c c c }
 \hline
 Method & Cost-aware & Quality-aware & Intra-session\\
 \hline
 Industry\cite{unreeling} & \ding{51} & \ding{51} & \ding{55} \\
 CFA~\cite{jiang2016cfa} & \ding{55} & \ding{51} & \ding{55} \\
 DIG~\cite{meng2022dig} & \ding{55} & \ding{51} & \ding{55} \\
 Ours & \ding{51} & \ding{51} & \ding{51} \\
 \hline\label{tab:methodcompare}
\end{tabular}
\end{center}
\vspace{-2em}
\end{table}

To evaluate PIRA’s performance and efficiency, we implemented a short video playback simulator with a multi-pan-CDN emulation module, enabling pan-CDN selection experiments based on real network throughput traces.
Simulation results show that, compared to video-level pan-CDN selection, PIRA reduces cost by 24\%, improves QoE by 32\%, while maintaining high runtime efficiency.
We further implement PIRA in our production video platform and conduct $A/B$ testing involving $450,000$ users over two weeks.
The experiment result shows that start-up delay is reduced by $2.1\%$, rebuffering occurrences decrease by $13.4\%$, and rebuffering duration is shortened by $15.2\%$.
Concurrently, PIRA achieves a $10\%$ reduction in unit traffic cost, showcasing its effectiveness in balancing user experience and operational efficiency at scale. Our contributions in this work are summarized as follows:
\begin{itemize}[leftmargin=0pt, itemindent=1em]
    \item We propose a media-list-level modeling framework for QoE and traffic cost, formally characterizing the trade-off between user experience and operational expenses in short-video streaming.
    \item We propose PIRA, an intra-video pan-CDN adaptation framework designed to address network dynamics in short-video streaming.
    Specifically, we introduce a control-theoretic approach that jointly optimizes pan-CDN selection and video segment duration, offering theoretical guarantees for balancing QoE and traffic cost.
    \item We implement PIRA on the client side and validate its effectiveness through large-scale production experiments, demonstrating substantial improvements in QoE and reductions in traffic costs compared to the baseline strategy.
\end{itemize}



\section{Motivation}\label{background}
\subsection{Cost and performance diversity of pan-CDN}
\begin{figure}[t]
\centering
    \begin{minipage}{0.49\linewidth}
        \centering
        \includegraphics[width=1\textwidth]{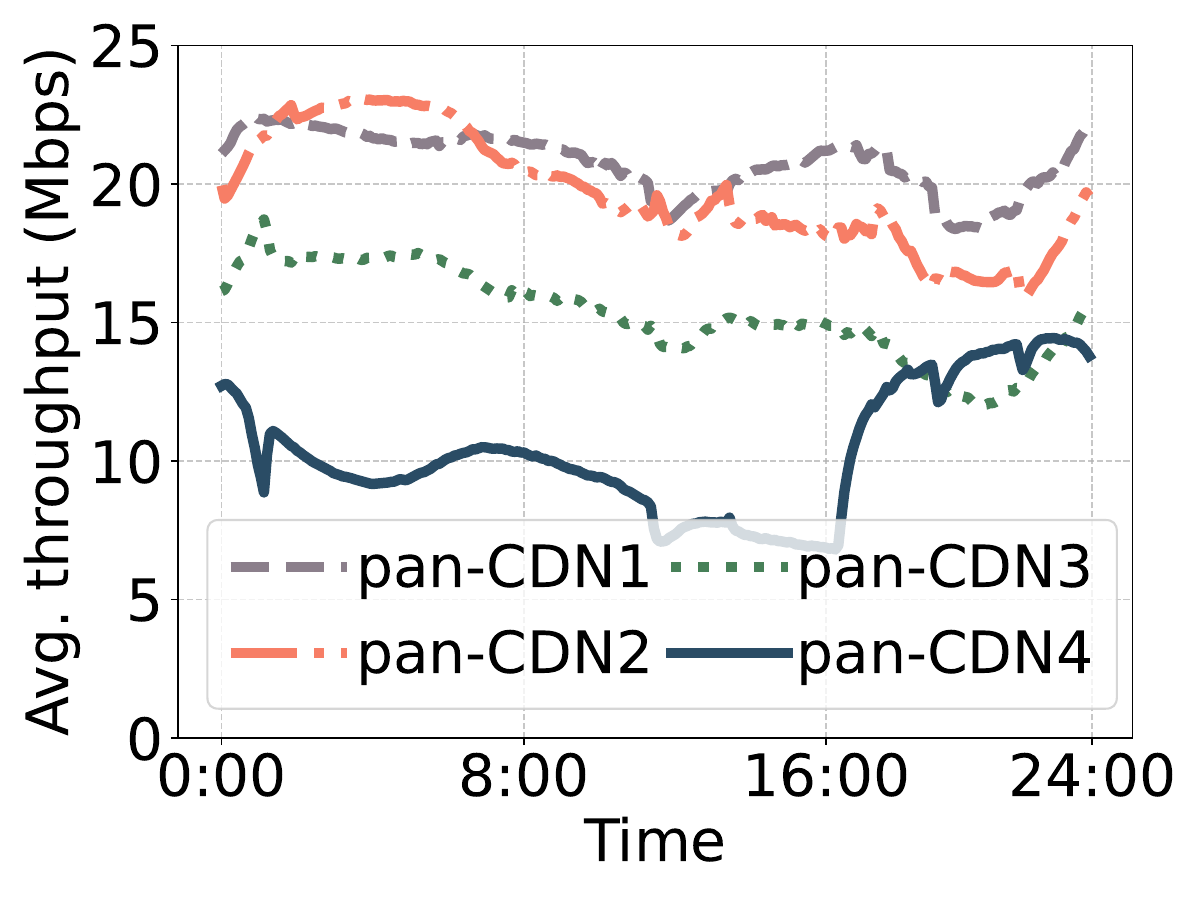}
        \centering{\footnotesize{(a)}}
    \end{minipage}
    \begin{minipage}{0.49\linewidth}
        \centering
        \includegraphics[width=1\textwidth]{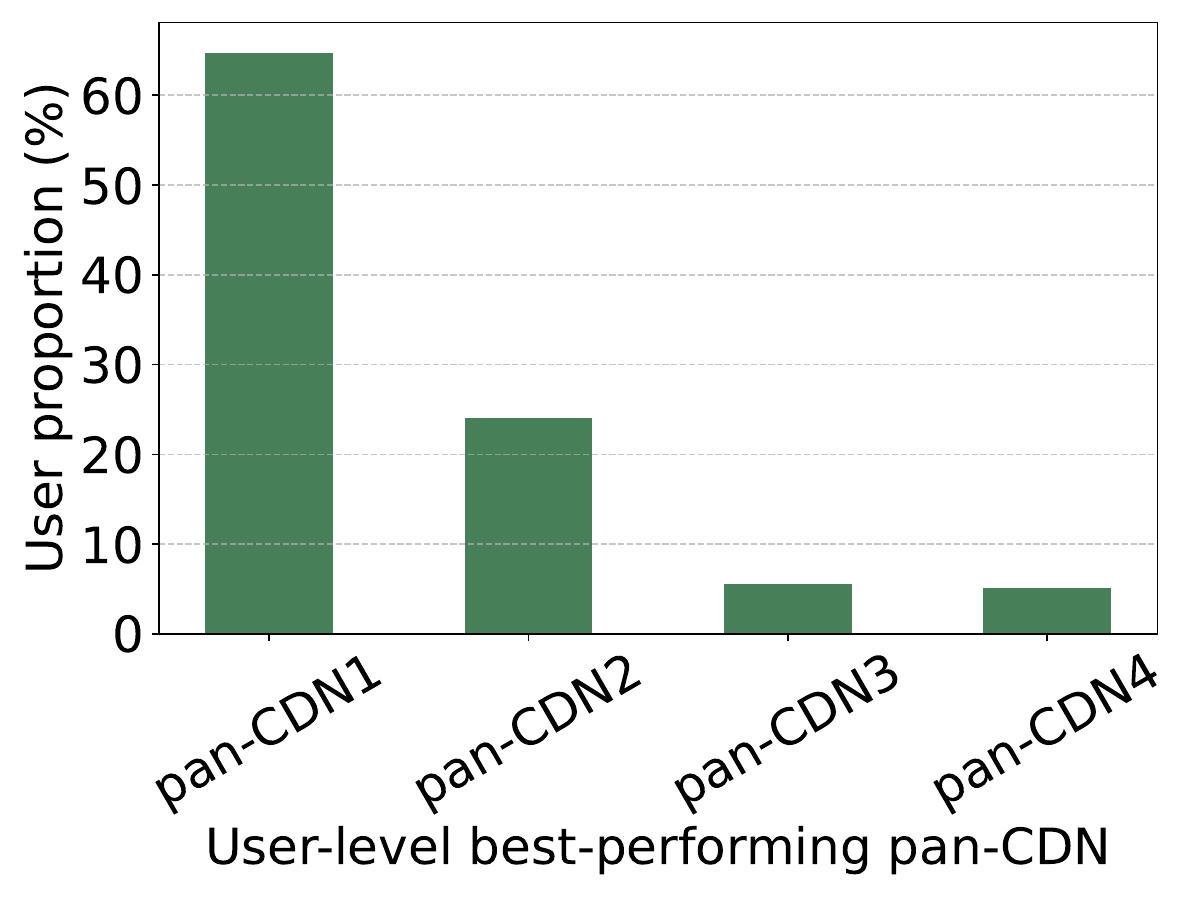}
        \centering{\footnotesize{(b)}}
    \end{minipage}
    \caption{Figure (a) illustrates the average throughput of pan-CDNs across all users on the video platform, while Figure (b) shows the proportion of users who experience the best performance from each pan-CDN. For instance, pan-CDN1 delivers the highest average bandwidth for $65\%$ of users.}
    \label{fig:stats}
    \vspace{-7pt}
\end{figure}

To characterize pan-CDN performance dynamics, we measure network download throughput across four pan-CDN types in our video production platform, Douyin, collecting video viewing traces from $1$ million mobile users.
Throughput is computed as the ratio of downloaded segment size to download time for each video segment.
To analyze temporal variations, we aggregate average throughput into 5-minute time slots across a 24-hour period.
As shown in Figure~\ref{fig:stats} (a), the average download speeds of pan-CDN $1$ to $4$ are $25$, $22$, $18$, and $14$ Mbps, respectively, reflecting their cost hierarchy.
However, daily throughput profiles exhibit significant temporal divergence: pan-CDN3 outperforms pan-CDN4 during most hours but is overtaken by pan-CDN4 during evening peak times.
A similar trend is observed between pan-CDN1 and pan-CDN2 during late-night low-traffic periods.

From a user-centric perspective, we rank pan-CDNs based on the average throughput per user, focusing specifically on users with over $1,000$ download records to ensure adequate utilization of all pan-CDN resources.
Figure~\ref{fig:stats} (b) illustrates the distribution of top-performing pan-CDNs among these users: approximately $65\%$ experience the highest average daily throughput from pan-CDN1, while $23\%$, $6\%$, and $6\%$ favor pan-CDN2, pan-CDN3, and pan-CDN4, respectively.
These results suggest that, despite its premium cost, pan-CDN1 does not consistently deliver the best performance for all users. 
We attribute this observation to heterogeneous network topologies. Pan-CDN resources are widely distributed, meaning the network topology between a user and a pan-CDN can vary significantly. For instance, edge nodes of a pan-CDN may be closer to the user than traditional CDN servers~\cite{ye2024kepc,zhang2024enhancing}.

Based on real-world measurement, we observe that although pan-CDN cost generally correlates with overall performance, the actual download throughput experienced by users exhibits far greater variability than expected.
These findings underscore the necessity of investigating two critical aspects: (1) identifying the factors responsible for the observed discrepancy between pan-CDN cost and user-experienced throughput, and (2) exploring methods for dynamically leveraging heterogeneous pan-CDN resources to simultaneously maximize QoE and optimize traffic costs.

\subsection{Intra-video pan-CDN adaptation}\label{intra-adapt}
\begin{figure}[t]
\centering
    \begin{minipage}{.75\linewidth}
        \centering
        \includegraphics[width=\textwidth]{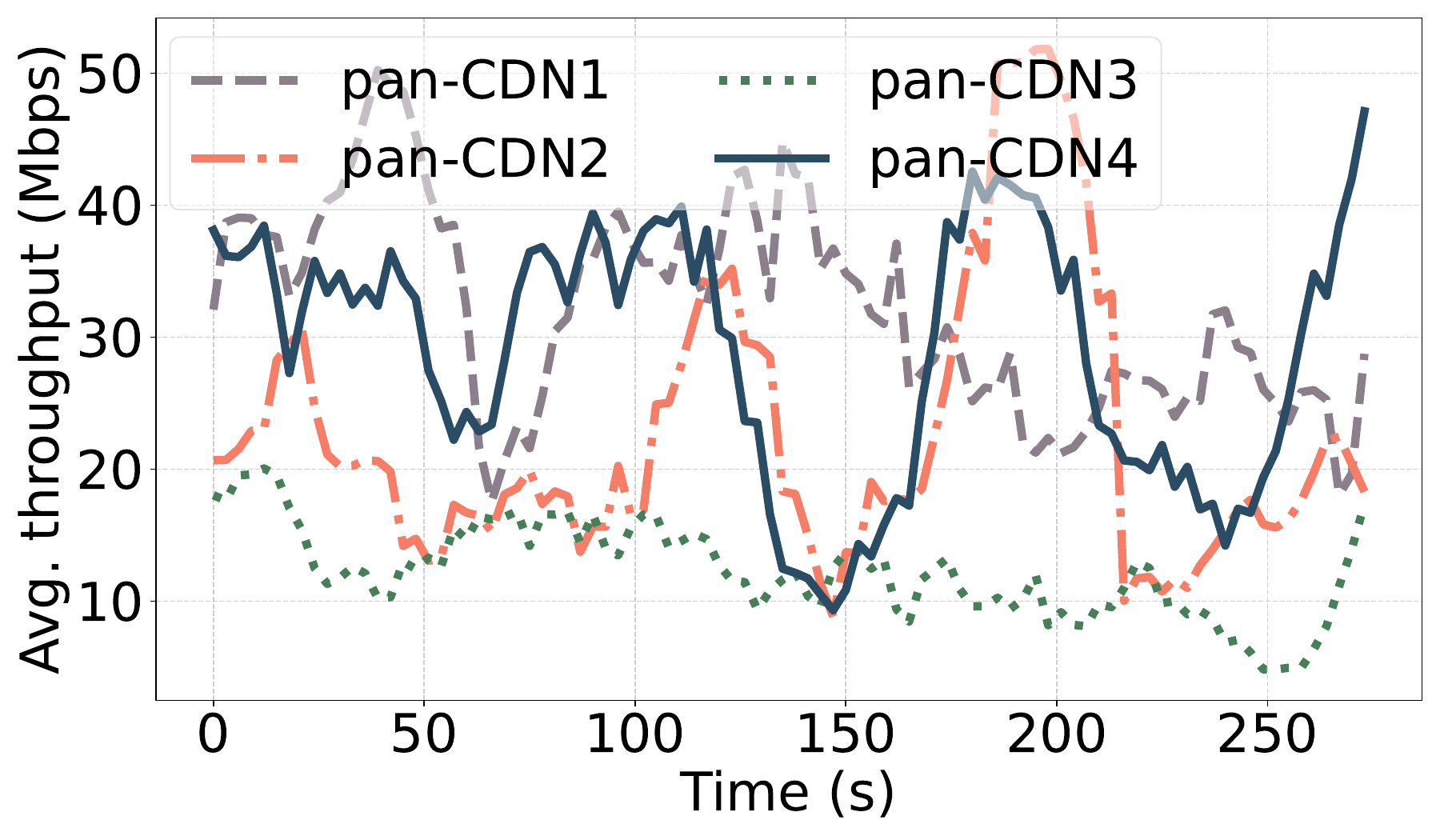}
    \end{minipage}
    \caption{An example of network throughput traces for different pan-CDNs in a short time period.}
    \label{fig:shortthroughput}
\end{figure}

We conducted measurements of per-segment download throughput using randomly assigned pan-CDNs.
As shown in Figure~\ref{fig:shortthroughput}, the results reveal significant fluctuations in throughput within seconds, highlighting the volatility of network conditions during video sessions.
This issue is even more pronounced on short video platforms. As illustrated in Figure~\ref{fig:medialist}, in short video streaming, clients download one video segment at a time from the media list.
To enhance video swipe interactions, upcoming video segments may be pre-downloaded before the current one finishes.
Consequently, a single pan-CDN may not remain optimal for the entire video download process.

We identify the following factors contributing to the variability in pan-CDN performance:
\begin{itemize}[leftmargin=0pt, itemindent=1em]
\item \textbf{Network link variability.} The network paths from different pan-CDN sources to end-user devices inherently exhibit temporal fluctuations in throughput and latency, due to factors such as congestion, routing dynamics, and access network conditions.

\item \textbf{Resource supply-demand dynamics.} Performance heavily depends on the real-time balance between user download requests and pan-CDN capacity.
When capacity exceeds demand, optimal throughput is achieved. However, during traffic spikes, resource contention can degrade performance, especially in low-cost pan-CDNs with limited capacity reserves.

\item \textbf{Source variety in pan-CDN.} For a given pan-CDN type, the actual edge nodes used for connections vary depending on the provider, resulting in practical throughput fluctuations even within the same pan-CDN type.

\end{itemize}

The significant variability in pan-CDN performance and cost necessitates a dynamic framework to balance Quality of Experience (QoE) and traffic costs.
Existing studies on network resource selection~\cite{jiang2016cfa,meng2022dig} typically focus on server-side quality prediction and pre-session resource allocation, which generally outperform random assignment.
While industrial video platforms often use static, session-level pan-CDN selection.
Clients typically choose a resource type based on initial network conditions and maintain this choice throughout the session, switching to premium resources like pan-CDN1 only during network failures~\cite{personalized}.
These static strategies fail to adapt to intra-session dynamics, as they don't account for real-time throughput changes or evolving user behavior (e.g., video priorities during short-video swiping).
This lack of adaptability leads to suboptimal utilization of pan-CDN resources, compromising both QoE and traffic efficiency in short video streaming.

\section{Modeling}\label{modeling2}

\subsection{Short video streaming model}
\begin{figure}[t]
  \centering
  \includegraphics[width=1.00\linewidth]{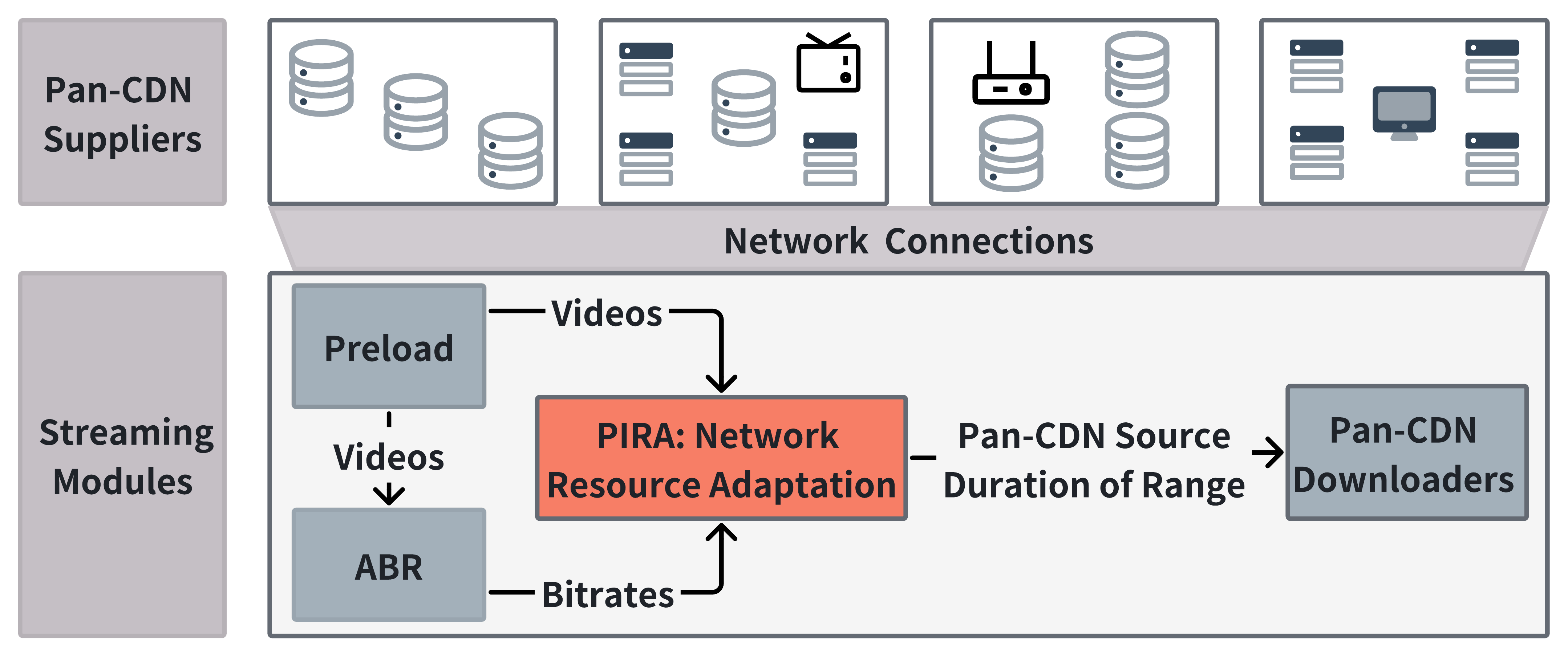}
  \caption{Overview of PIRA and associated client-side streaming modules.}
  \label{fig_framework}
  \vspace{-3pt}
\end{figure}

The video platform employs recommendation algorithms to generate personalized playlists and enables users to easily swipe content they are not interested in.
On the client side, multiple components collaboratively optimize video downloading, as illustrated in Figure~\ref{fig_framework}.
Firstly, to minimize rebuffering during swiping, adaptive preload algorithms set the download priorities, which predict user engagement to optimize the prefetching order.
Once a video is scheduled for download, an adaptive bitrate (ABR) algorithm selects the chunk bitrate.
PIRA then dynamically chooses the pan-CDN type and video segment duration based on real-time network conditions and the pre-defined QoE-cost tradeoff.
Finally, the client connects to the selected network resource and retrieves the video data.


The video platform utilizes a large array of edge devices from multiple pan-CDN vendors.
These devices vary significantly in hardware, transport protocols, and serving modes, resulting in notable variability in service quality and traffic costs.
To manage this complexity, network resources are classified into distinct pan-CDN types based on pricing tiers, streamlining the management of quality-cost tradeoffs across heterogeneous edge environments.

\begin{figure}[t]
  \centering
  \includegraphics[width=.8\linewidth]{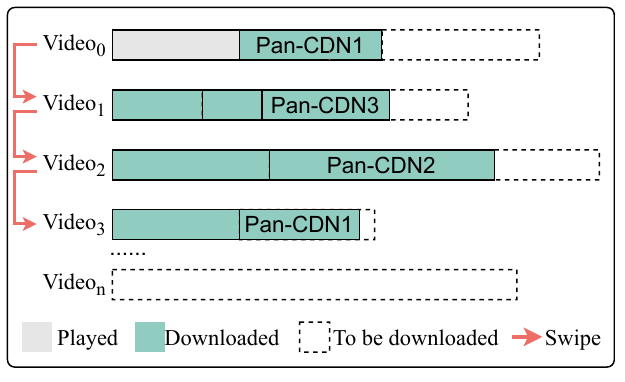}
  \caption{An illustrative example of the media-list download process in short-video streaming.}
  \label{fig:medialist}
  \vspace{-0pt}
\end{figure}

Short-video streaming follows a segmented download process, as shown in Figure~\ref{fig:medialist}.
When a user accesses the platform, they receive a media list $v_i$, where$\{i=1,2,3,...,n\}$ and $n$ denotes the number of videos in the list.
For each video $v_i$ scheduled for download, the system selects a pan-CDN type from the set $pc_{j}$, $\{j = 1, 2, 3, 4\}$, with $c(pc_j)$ representing the unit traffic cost coefficient for pan-CDN
$pc_j$.
Videos are downloaded in a segmented manner.
For example, a $4$-second video chunk encoded at $2$ Mbps may be further partitioned into $R_i$ sequential download ranges, the duration of a video range is ${r_k}$, where $\{k = 1, 2, 3, ..., R_i\}$. $r_k$ is no longer than the chunk length (e.g., 4 seconds, in line with DASH protocol standards~\cite{lederer2012dynamic}).
This partitioning strategy allows for dynamic pan-CDN switching within a chunk, optimizing the balance between streaming quality and bandwidth consumption by leveraging real-time network conditions across download ranges.

The playback buffer evolves dynamically during both video streaming and progressive prefetching, influenced by user interactions (e.g., swiping between videos in the media list).
Rebuffering occurs when a user swipes to a new video that has not been pre-downloaded, disrupting the viewing experience.
To model this, we define the buffer state for each video $v_i$ in the media list: let  $B(t_{k})|v_i$ denote the available buffer size (in seconds) for video $v_i$ at start time $t_k$ of download range $r_k$.
If the video $v_i$ to be downloaded is the same to the video the user is currently viewing, we have
\begin{equation}
B(t_{k+1})|v_i, pc_j = ((B(t_{k})|v_i - \frac{d_k(r_k)|v_i}{\overline{Th_k}(pc_j)})_{+} + r_k - \Delta t_{k})_+
\label{eqn:buffer_definition}
\end{equation}
where $d_k(r_k)|v_i$ denotes the size of video $v_i$ for range $r_k$, and $\overline{Th_k}(pc_j)$ denotes the average download throughput for range $r_k$ with pan-CDN $pc_j$.
The notation $(x)_+ = max(x, 0)$ assures that the variable (video buffer) is not negative.
All video buffer shares the same video player buffer $B_{vp}$ and the video player will wait to download range $r_k$ if the buffer is fulfilled by all videos' buffer.
The waiting time $\Delta t_{k}$ is defined as
\begin{equation}
 \Delta t_{k}  = ((B(t_{k})|v_i - \frac{d_k(r_k)|v_i}{\overline{Th_k}(pc_j)})_+ + \sum_{v_i \not \subset v}{B(t_k)|v} + r_k - B_{vp}^{max})_+
 \label{eqn:buffer_overflow}
\end{equation}

While if the download video $v_i$ is the next videos to the video $v_0$ of the user is currently viewing, the buffer can be formulated as
\begin{subequations}
 \begin{align}
    B(t_{k+1})|v_i &= (B(t_{k})|v_i + r_k - \Delta t_{k})_+\label{eqn:buffer_definition_2}\\
    B(t_{k+1})|v_0 &= (B(t_{k})|v_0  - \frac{d_k(r_k)|v_i}{\overline{Th_k}(pc_j)})_+\label{eqn:buffer_definition_3}
 \end{align}
\end{subequations}
where $v_{i} \neq v_{0}$.

\subsection{Problem formulation}
QoE is typically modeled using multi-dimensional metrics, including video quality, rebuffering, video quality switching, and startup delay in ABR algorithms~\cite{mok2011inferring,yin2015control,akhtar2018oboe,qiao2020beyond,qiao2021understanding,zhang2023quty}.
Since our paper focuses on dynamic network resource selection rather than bitrate optimization, we narrow our QoE model to rebuffering rate/duration and startup delay.

Rebuffering occurs when the playback buffer of the currently viewed video is depleted, i.e., when the available buffer duration drops to zero during playback.
To mitigate this, the preload algorithm dynamically triggers the pre-download of the current video when the buffer level falls below a predefined threshold, balancing proactive resource allocation with real-time buffer dynamics.
Thus the rebuffering time for downloading range $r_k$ of $v_i$ (a user is viewing) can be calculated by
\begin{equation}
  T_{r_k} =  (\frac{d_k(r_k)|v_i}{\overline{Th_k}(pc_j)} - B(t_{k})|v_i)_+ \label{eqn:rebuf1}
\end{equation}

The start-up delay of a video is defined as the time interval between the user's initiation of playback and the rendering of the first frame.
We introduce a constant variable $\tau_{st}$ representing the minimum buffer duration required to trigger playback initialization (e.g., 2 seconds of buffered content).
The start-up delay $T_s(v_i)$ for video $v_i$ is formally defined as
\begin{equation}
T_s(v_i) = \begin{cases}
    0, & \text{ when } B(t_{k})|v_i \geq \tau_{st}, \\
    \frac{d_k(r_k)|v_i}{\overline{Th_k}(pc_j)}, & \text{ when } B(t_{k})|v_i < \tau_{st}
 \end{cases}\label{eqn:st2}
\end{equation}
where $r_k + B(t_{k})|v_i \geq \tau_{st}$.
Equation~\ref{eqn:st2} implies when the video startup initialization fails due to insufficient buffer levels, the video player prioritizes data downloading for video $v_i$.

The traffic cost of download video data is
\begin{equation}
    Cost = \sum_{i=1}^{n}\sum_{k=1}^{R_i}d_k(r_k)|v_i * c(pc_{r_k})
    \label{eqn:cost1}
\end{equation}
for videos $v_i$, $i = \{i=1,2,3,...,N\}$.
$pc_{r_k}$ denotes the pan-CDN type for downloading range $r_k$.

We define the QoE function for video $v_i$ as
\begin{equation}
    QoE_{v_i} = 1 - \mu_1\frac{\sum_{k=1}^{R_i}{T_{r_k}}}{T_{v_i}}  - \mu_2T_s(v_i)
    \label{eqn:qoe1}
\end{equation}
where $\mu_1, \mu_2 > 0$.
$T_{v_i}$ denotes the watch duration for video $v_i$.
Parameter $\mu_1$ denotes the penalty weight on rebuffering ratio and $\mu_2$ denotes the weight on start-up delay of video $v_i$.

As shown in Figure~\ref{fig:medialist}, the selection of pan-CDNs has cascading effects.
For example, when the video player selects an inappropriate pan-CDN, it may induce a longer download time, and the next videos are also impacted.
To this end, we conduct the media list level QoE definition based on Equation~\ref{eqn:qoe1} as follows:
\begin{equation}
    QoE = \sum_{i=1}^{n}QoE_{v_i}
    \label{eqn:qoe2}
\end{equation}

Our paper aims to design an pan-CDN selection algorithm to achieve the optimal trade off (utility) between QoE and traffic cost.
Formally,
\begin{align}
    \underset{pc,r}{\text{maximize}} \quad & {QoE - \gamma Cost}\ \label{eq:opt1} \\
    \text{subject to} \quad & \text{Equation ~\ref{eqn:buffer_definition},~\ref{eqn:buffer_overflow},~\ref{eqn:buffer_definition_2},~\ref{eqn:buffer_definition_3},~\ref{eqn:rebuf1},~\ref{eqn:st2}} \nonumber
\end{align}
where $\gamma$ is a non-negative weighting parameter for traffic cost over QoE.

\section{Design}\label{design}
\subsection{Design rationale}
As formulated in Equation~\ref{eq:opt1}, the joint pan-CDN and video range duration selection problem constitutes a stochastic optimal control problem.
However, its practical solution presents several challenges, detailed as follows:
\begin{itemize}[leftmargin=0pt, itemindent=1em]

\item Joint selection of pan-CDNs and video ranges affects system dynamics.
Low-throughput choices may drain buffers quickly, triggering costly switches to avoid rebuffering—revealing a trade-off between cost and responsiveness.
Shorter ranges enable agile switching, while longer ranges reduce overhead but limit adaptation to throughput changes.

\item Accurately predicting pan-CDN quality is difficult due to wireless network volatility.
Switching incurs connection overhead (e.g., TCP handshake, slow start~\cite{jacobson1988congestion,floyd1999promoting}), causing temporary throughput drops.
Limited historical data—since each segment uses only one pan-CDN—further hinders accurate estimation.

\end{itemize}


\begin{algorithm}[t]
    \caption{Workflow of PIRA}
    \footnotesize
    \begin{algorithmic}[1]
        \State $V =\{v_i\}$ \Comment{Get video download sequences}
        \State $PC = \{pc_j\}, R = \{r_k\}$  \Comment{Get alternative pan-CDNs and video ranges}
        \If {need to update video download sequences}
            \State  update $V$, go to 1
        \Else
            \State  $\overline{Th}$ = $\{\overline{Th_k}(pc_j), \overline{Th_{k+1}}(pc_j), ... \} = ThroughputPred(V, PC, R)$
            \State  $\{v_i, pc_j, r_k\} = f_{pira}^{n}(V, PC, R, \overline{Th})$
        \EndIf
        \State Download range $r_k$ with pan-CDN $pc_j$ for video $v_i$, wait till finished
        \State Update pan-CDNs' history throughput, probe if necessary
    \end{algorithmic}
    \label{alg:mpc}
\end{algorithm}
To address the challenges of pan-CDN switching costs and the cascading effects of joint resource–range decisions, we propose PIRA (Pan-CDN Intra-video Resource Adaptation), built on a MPC framework~\cite{mpc}, which enables forward-looking joint optimization of QoE and delivery cost over a receding time horizon.
At each decision epoch (aligned with the segment range $r_k$), PIRA optimizes over a short planning horizon, modeling network throughput dynamics and buffer evolution.
The control model incorporates penalties for pan-CDN switching and transmission delays, ensuring decisions balance immediate QoE needs with long-term traffic efficiency.
Using predictions of pan-CDN performance, PIRA selects the optimal resource-duration combination by evaluating expected QoE and traffic costs across all feasible future states.





\subsection{PIRA design}\label{sec:piradesign}
The workflow of PIRA is shown in Algorithm~\ref{alg:mpc}:
(1) Identify alternative pan-CDNs and video ranges based on the current video sequence.
(2) Check if the video sequence needs updating before downloading a new video range, and update if necessary.
(3) Predict the average network throughput for the next horizons across different pan-CDNs.
(4) Calculate the optimal pan-CDN type and the corresponding video range duration.


\subsubsection{Determining candidate pan-CDNs and video download range durations.}
Pan-CDN selection depends on video caching status, which follows two policies: real-time origin pull and delayed caching based on popularity.
For uncached videos, the system prefers real-time origin pull to reduce playback latency.
Cache states are tracked server-side and indicated in video playlists, enabling efficient CDN use and minimizing buffering risk.
To reduce the decision space, we constrain video range duration. It must not exceed the chunk length and should be long enough to avoid frequent pan-CDN switching and bandwidth underutilization.
As shown in Figure~\ref{fig:stats}, high throughput often removes the need for splitting.
For example, a 4-second chunk at 2 Mbps takes 1 second to download via the slowest pan-CDN, while splitting into 1-second segments only marginally improves latency.

\begin{algorithm}[t]
    \footnotesize
    \caption{Design of $f_{pira}^{n}(V, PC, R, \overline{Th})$.}
    \begin{algorithmic}[1]
        \Require Get download video, alternative pan-CDN and video range lengths, and predicted network throughput $V, PC, R, \overline{Th}$
        \Ensure pan-CDN $pc_j$ and range $r_{k+1}$ at time $t_{k+1}$ for the current video

        \State Initialize parameters, $\gamma, \mu_1, \mu_2, n$, rewards map $M = \{\}$
        \For {i = 1 to n}
            \State $S_{can}^i \gets$ \text{Get all pan-CDN and range candidates at step $i$}
                \State $S_{Pruning}^i \gets$ \text{Pruning}
            \For {$\{(pc_1, r_1), ..., (pc_i,r_i)\}$ in $S_{Pruning}^i$}
                \State calculate each step's reward according to Equation~\ref{eq:opt1}
                \State update $M$ of key $\{(pc_1, r_1), ..., (pc_i,r_i)\}$ with utility value of $QoE-\gamma Cost$
            \EndFor
        \EndFor
        \State select $(pc^*, r^*)$ that maximize $M$ as output
    \end{algorithmic}
    \label{alg:mpc_spec}
\end{algorithm}
\subsubsection{Network throughput prediction.}
The throughput prediction faces two key challenges: the heterogeneous performance of pan-CDNs over short time scales and the overhead introduced by pan-CDN switching.
Unlike generic approaches that treat networks as homogeneous, PIRA explicitly tracks historical throughput traces for each pan-CDN and employs independent prediction models to capture their unique temporal dynamics.
To account for throughput degradation due to pan-CDN switching, PIRA incorporates a degradation coefficient to model slow start effects from flow control and startup delays (e.g., 1.5 times average RTT) into its throughput prediction when evaluating pan-CDN transitions.
PIRA adopts a harmonic mean (HM)-based approach~\cite{yin2015control} for future throughput estimation, leveraging its simplicity and effectiveness in mitigating transient switching overheads.

Since only one pan-CDN is used per video download, some pan-CDNs may have limited historical throughput data due to infrequent use.
To address this, PIRA includes a dedicated module that proactively probes network throughput for each pan-CDN.
Probing intervals are set to tens of seconds, minimizing overhead while balancing the need for updated performance data and efficient resource utilization.

\subsubsection{Algorithm design.}

The function $f_{pira}^{n}(V,PC,R,\overline{Th})$ is the key algorithm of our design, where $n$ denotes  the look-ahead horizon.
As detailed in Algorithm~\ref{alg:mpc_spec}, PIRA enumerates all feasible future states within this planning horizon, considering candidate pan-CDNs $pc_j$ and discrete download range duration $r_k$.
For each state sequence, the algorithm evaluates QoE and traffic cost by integrating buffer evolution dynamics (defined in Equation~\ref{eqn:buffer_definition}) and throughput predictions for each pan-CDN.
After completing the $n$-step receding-horizon optimization, PIRA selects the optimal combination of pan-CDN $c^*$ and video range duration $r_k^*$ that maximizes Equation~\ref{eq:opt1}.
This systematic state-space enumeration explicitly captures interdependencies between resource selection, range granularity, and buffer status, ensuring adaptability to dynamic network conditions while respecting real-time edge computation constraints.

   \begin{figure}[t]
   \vspace{-8pt}
    \centering
        \begin{minipage}{.85\linewidth}
            \centering
            \includegraphics[width=\textwidth]{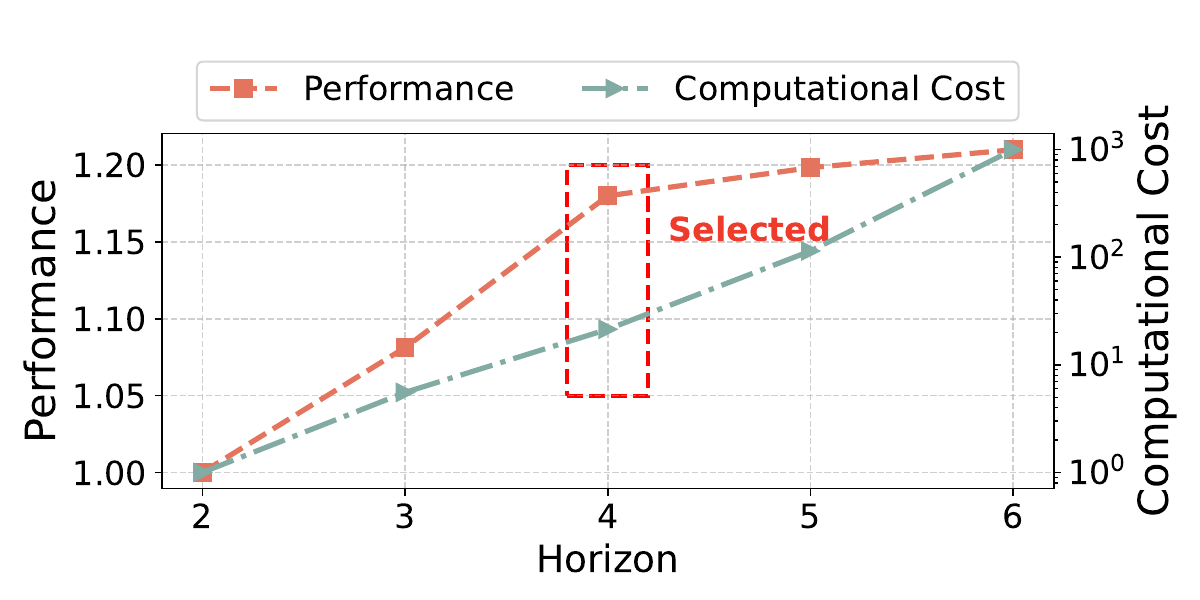}
        \end{minipage}
    \vspace{-2pt}
        \caption{Normalized performance and computation overheads of different planning horizon steps in PIRA. Performance $=$ QoE-$\gamma$Cost, where $\gamma = 0.3$.}
        \label{exp:horizon}
        \vspace{-5pt}
    \end{figure}
\begin{figure}[h]
    \centering
        \begin{minipage}{.75\linewidth}
            \centering
            \includegraphics[width=\textwidth]{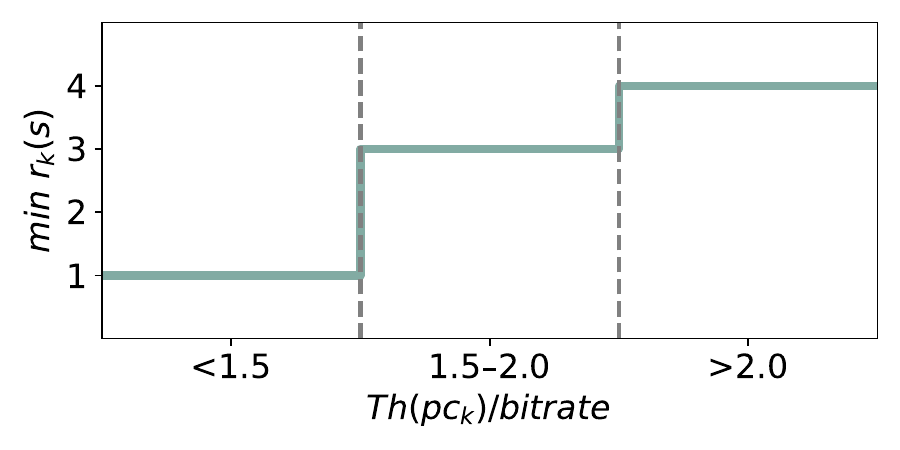}
        \end{minipage}
         \vspace{-2pt}
        \caption{
        For pan-CDN$k$, PIRA adjusts the exploration space for download duration dynamically, based on the ratio between its predicted throughput and the video bitrate.}
        \label{exp:rk}
        \vspace{-5pt}
    \end{figure}

The computational complexity in Algorithm~\ref{alg:mpc_spec} is $(len(PC) * len(R)) + (len(PC) * len(R))^2 + ... + (len(PC) * len(R))^n$.
With 4 pan-CDNs and video chunks discretized into segments ranging from 1 to 4 seconds, the time complexity of Algorithm~\ref{alg:mpc_spec} is $O(2^{4n})$, which grows exponentially with the planning horizon length $n$.
To reduce computation time, prior work such as~\cite{yin2015control} uses indexing tables to store precomputed optimal results for each system state.
However, the significant memory overhead makes it impractical for real-world deployment.
To make PIRA a high-performance and deployable solution, we propose two customized pruning strategies tailored to the pan-CDN selection scenario:
\begin{itemize}[leftmargin=0pt, itemindent=1em]

    \item \textbf{Pruning \uppercase\expandafter{\romannumeral1}: Pan-CDN resource filtering.} A clear pricing hierarchy exists among candidate pan-CDNs, though their throughput may fluctuate over time.  As demonstrated in Section~\ref{background}, higher cost does not always guarantee better performance.
    Therefore, during the exploration phase, PIRA excludes pan-CDNs that exhibit both lower throughput and higher costs compared to available alternatives.
    Let $Th_i$ and $c_i$ denote the estimated throughput and cost of candidate pan-CDN $i \in S_{\text{can}}^i$.
    Then, the pruned candidate set $S_{\text{Pruning\uppercase\expandafter{\romannumeral1}}}^i$ is defined as:
    \begin{align}
    S_{\text{Pruning\uppercase\expandafter{\romannumeral1}}}^i = \Big\{
    & (pc_k, r_k) \in S_{\text{can}}^i \,\Big|\, \nexists (pc_j, r_j) \in S_{\text{can}}^i,\ j \ne k, \notag \\
    & \text{s.t. } Th(pc_j) > Th(pc_k)\ \text{and } c(pc_j) < c(pc_k)
    \Big\}
    \end{align}

    \item \textbf{Pruning \uppercase\expandafter{\romannumeral2}: Range duration filtering.}
    When network conditions are poor, it's better to select expensive, high-throughput resources and explore shorter durations to retain flexibility for switching to cheaper alternatives later.
    Conversely, when the estimated throughput of a pan-CDN significantly exceeds the video bitrate, or the playback buffer is large enough to prevent rebuffering, shorter durations offer little benefit and only add unnecessary request overhead.
    Therefore, we construct a step-wise mapping function to prune PIRA’s search space more effectively.
    As shown in Figure~\ref{exp:rk}, the minimum download duration  $r_k$ is constrained by the ratio of expected future throughput to video bitrate.


\end{itemize}

\begin{figure*}[t]
    \begin{minipage}[b]{1\textwidth}
        \centering
        \includegraphics[width=0.73\linewidth]{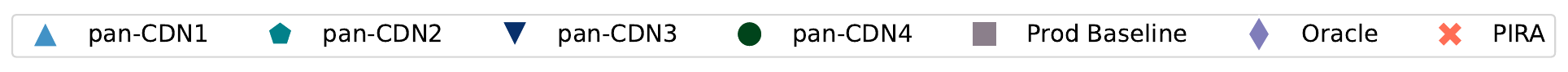}
        \begin{minipage}[b]{0.3\textwidth}
            \centering
            \includegraphics[width=\linewidth]{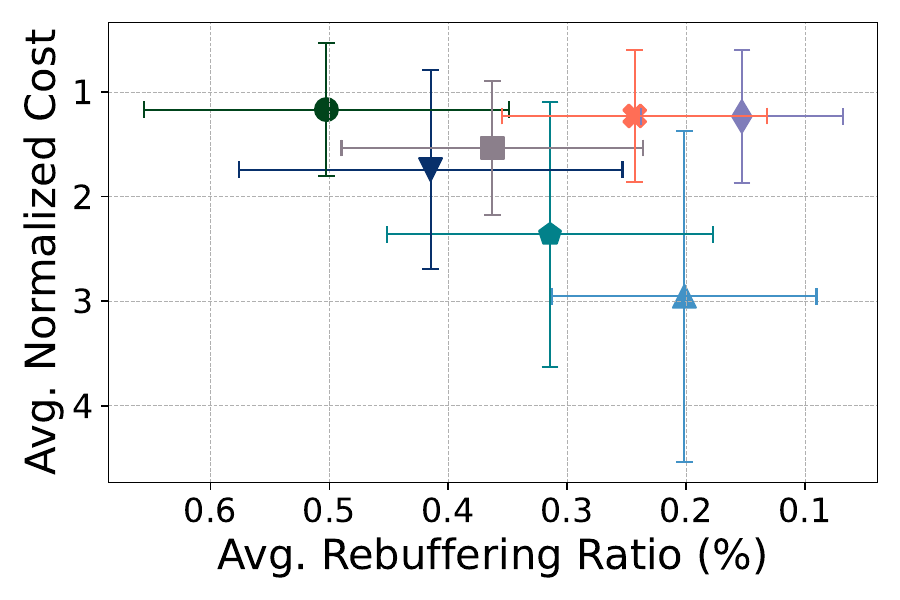}
            \centering{{(a) 0:00-8:00}}
            \label{fig:lab}
        \end{minipage}%
        \hfill
        \begin{minipage}[b]{0.3\textwidth}
            \centering
            \includegraphics[width=\linewidth]{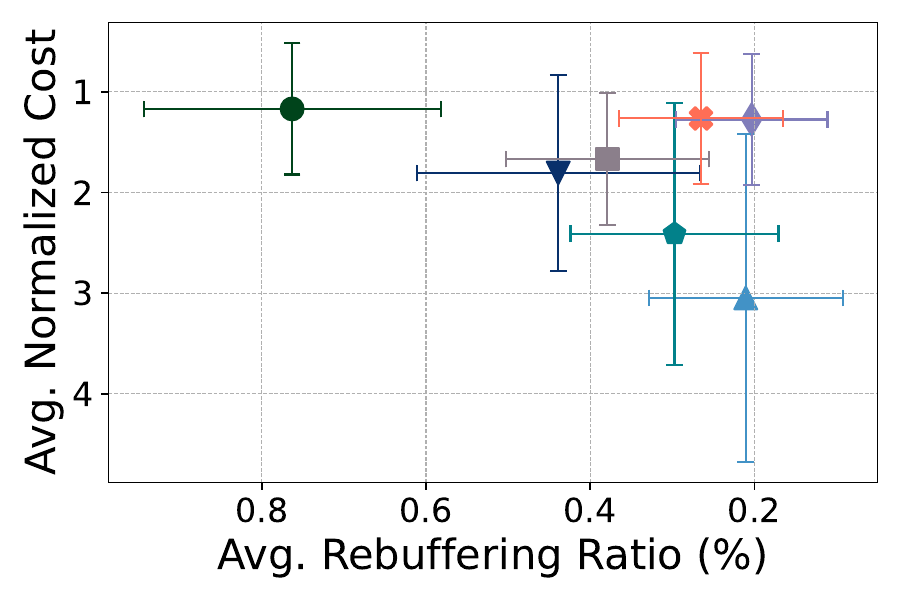}
            \centering{{(b) 9:00-18:00}}
            \label{fig:bus}
        \end{minipage}%
        \hfill
        \begin{minipage}[b]{0.3\textwidth}
            \centering
            \includegraphics[width=\linewidth]{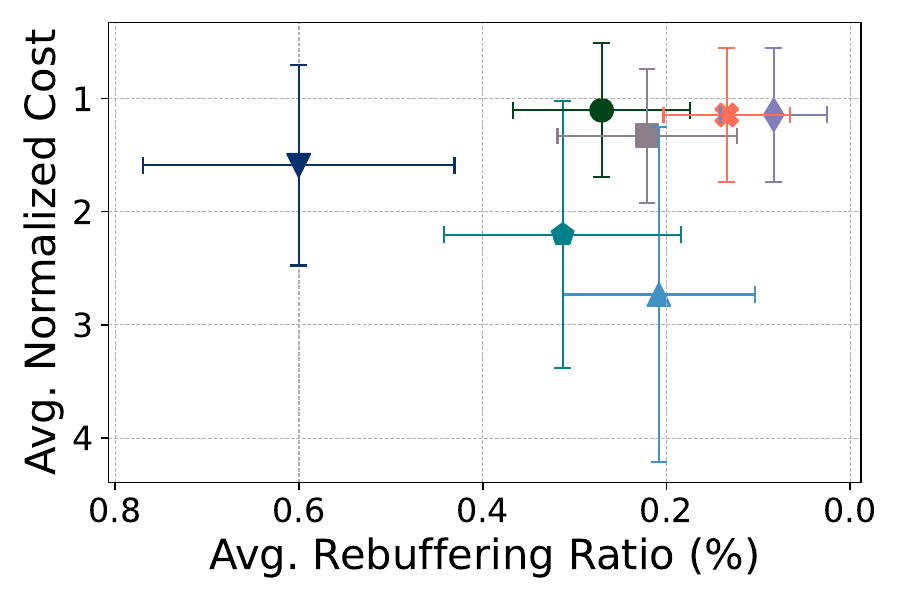}
            \centering{{(c) 19:00-24:00}}
            \label{fig:train}
        \end{minipage}%
        \caption{Simulation results comparing PIRA with baseline methods are presented, highlighting rebuffering rate and normalized traffic cost metrics. Error bars indicate $95\%$ confidence intervals.}\label{exp:v1}
    \end{minipage}%
    \vspace{-5pt}
\end{figure*}
In our scenario, the planning horizon $n$ can remain relatively small.
For example, when a user swipes through short videos, the video download sequence in Algorithm~\ref{alg:mpc} will adjust accordingly.
The median viewing time of each video is typically less than 15 seconds on platforms~\cite{li2023dashlet}.
Additionally, while network conditions fluctuate over time, they remain reasonably stable on short timescales (tens of seconds)~\cite{zhang2001constancy}.
As shown in Figure~\ref{exp:horizon}, results demonstrate that when $n = 4$ (considering the next four video ranges), PIRA achieves an optimal tradeoff between performance and computational efficiency.

\section{Evaluation}\label{evaluation}

To assess PIRA’s performance, we conduct two types of experiments: 1) simulation experiments to validate its performance and efficiency, and 2) a production deployment on our short-video platform, Douyin, to evaluate real-world applicability.
As outlined in previous sections, PIRA’s default parameters are set as $n = 4$ (planning horizon steps) and $\gamma = 0.3$ (QoE-cost tradeoff coefficient).


\subsection{Numerical simulations}


To validate PIRA’s performance in a controlled environment, we implemented a pan-CDN selection simulator, adapted from an ABR simulator~\cite{mao2017neural} with verified accuracy.
The simulator concurrently models network links from four different pan-CDNs (labeled pan-CDN1 to pan-CDN4) and simulates the logic for video downloading and playback.

\begin{figure}[t]
\centering
\begin{minipage}{.85\linewidth}
    \centering
    \includegraphics[width=\textwidth]{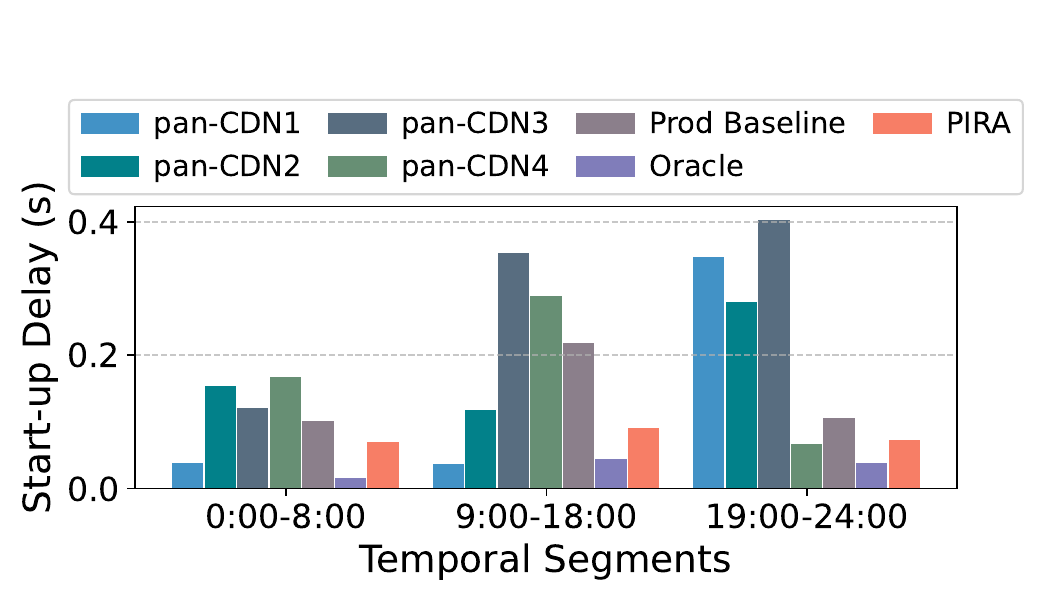}
    \caption{Simulation results comparing PIRA with baseline methods in terms of start-up delay.}
    \label{exp:sd}
\end{minipage}


\begin{minipage}{.85\linewidth}
    \centering
    \includegraphics[width=\textwidth]{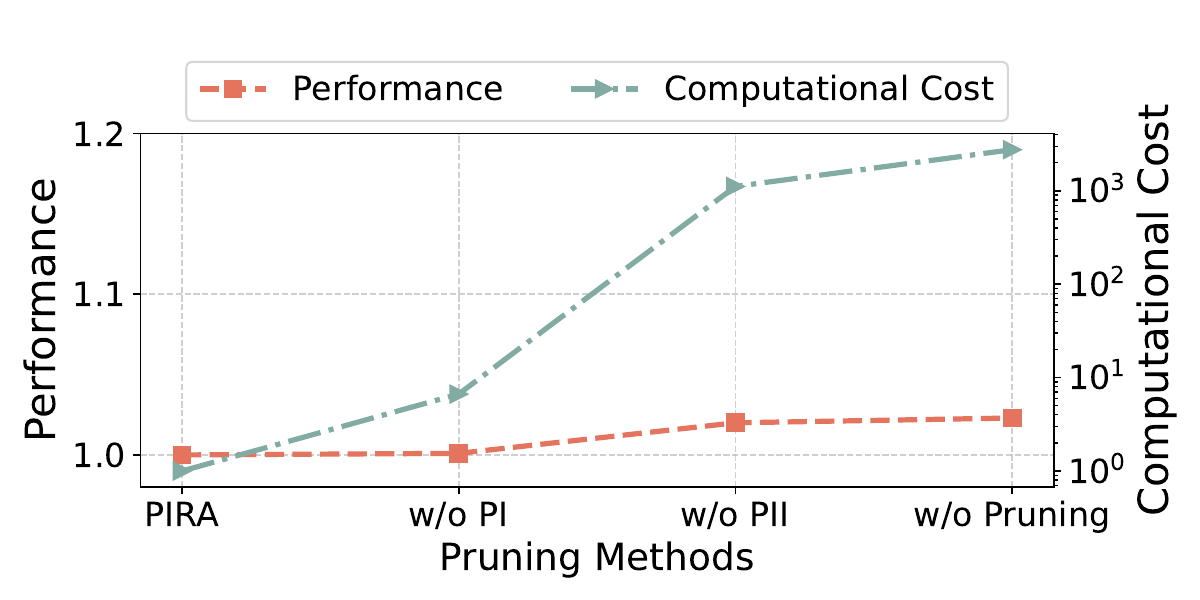}
    \caption{Normalized performance and computational overhead for different pruning methods.}
    \label{exp:pruning}
\end{minipage}\vspace{-5pt}
\end{figure}

\subsubsection{Simulation dataset.} Our network bandwidth dataset and video dataset were collected from the short video platform.
\begin{itemize}[leftmargin=0pt, itemindent=1em]
    \item Network data collection. We collected bandwidth data from different pan-CDNs during three time periods: $0-8$ (off-peak), $9-18$ (peak), and $19-24$ (evening peak).
    Due to technical constraints preventing simultaneous data collection from heterogeneous resources, we sequentially aggregated the bandwidth traces into a unified dataset, preserving temporal order to reflect real-world network dynamics.
    \item Video List Data. We gathered video playlists from the viewing history of 1,200 users, covering 2,300 hours of content, with $73\%$ of the videos having a duration of less than 30 seconds.
\end{itemize}

\subsubsection{Baselines.} We compare PIRA with the following client-side based methods: (1) \textbf{Pure pan-CDN:} Exclusive use of a single pan-CDN for all video deliveries, with no dynamic resource switching. (2) \textbf{Production baseline:} The platform’s deployed strategy, which selects the cheapest pan-CDN whose network throughput slightly exceeds the video bitrate to minimize buffer delay.
In case of rebuffering, it switches to pan-CDN1 (a high-reliability CDN) to resume downloading.
(3) \textbf{Oracle:} An idealized version of PIRA with full knowledge of future pan-CDN throughputs, serving as the performance upper bound.


\begin{figure*}[t]
\centering
    \begin{minipage}{.32\linewidth}
        \centering
        \includegraphics[width=\textwidth]{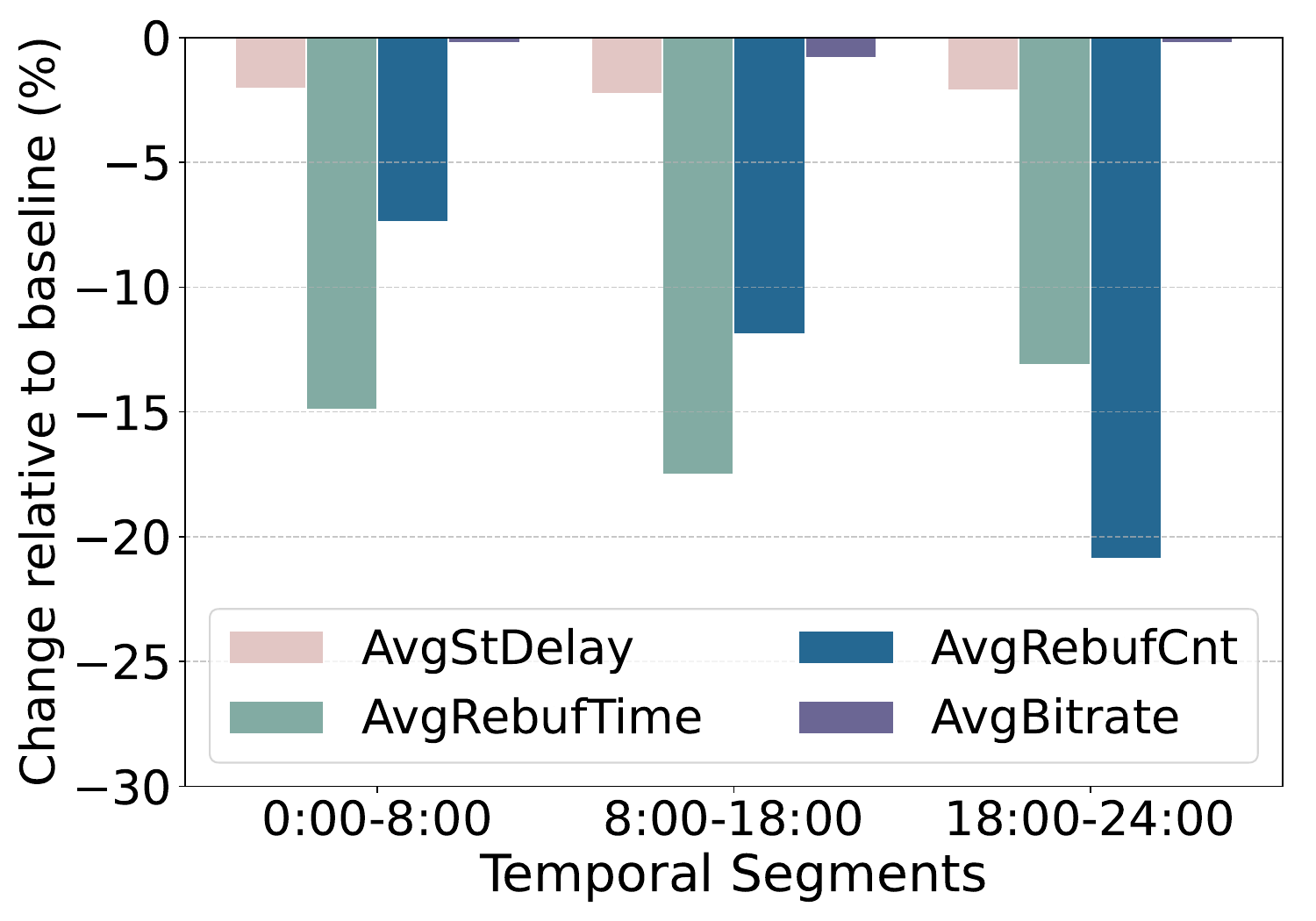}
        \centering{\footnotesize{(a) PIRA's performance on QoE metrics.}}
    \end{minipage}
    \begin{minipage}{.32\linewidth}
        \centering
        \includegraphics[width=\textwidth]{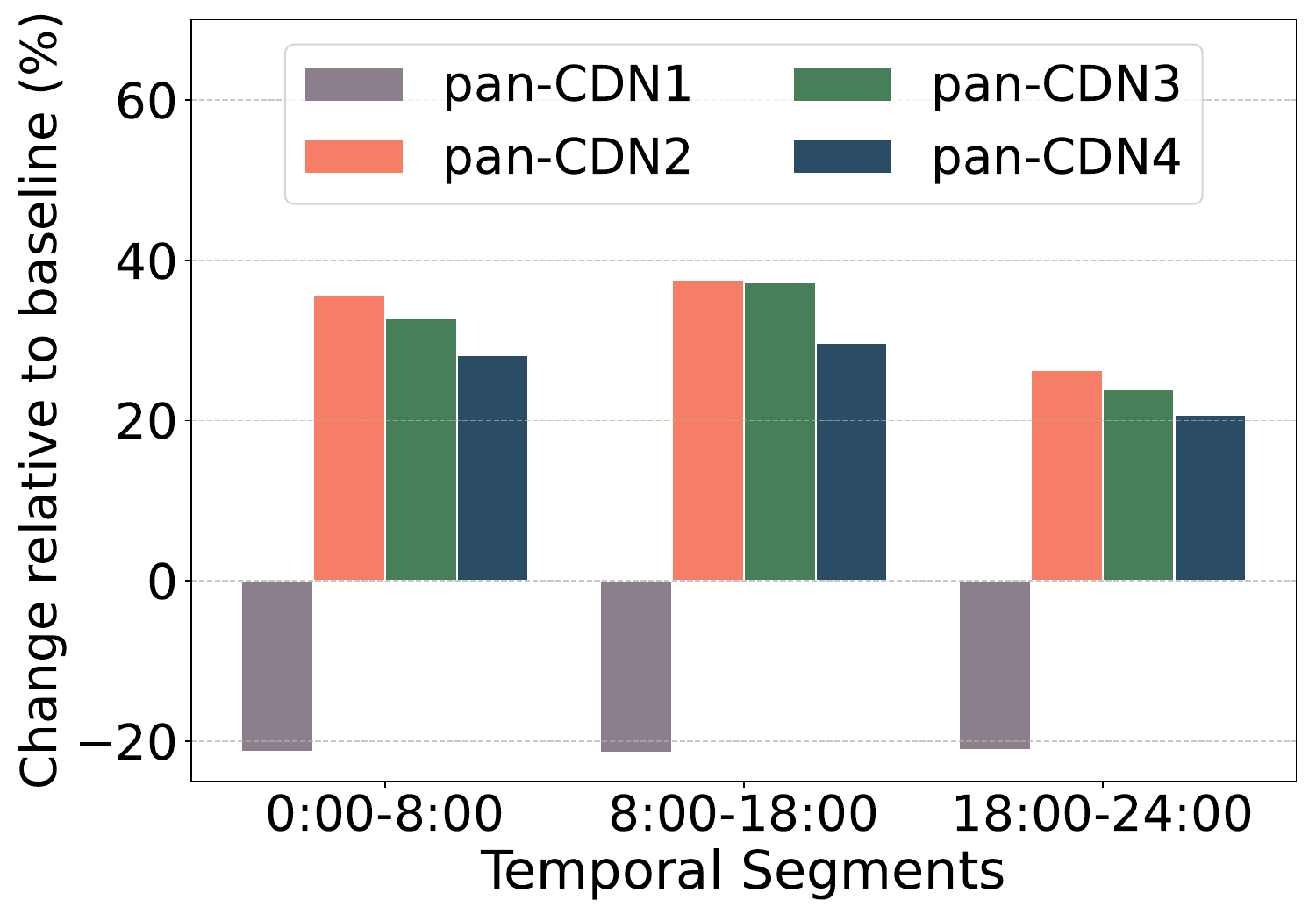}
        \centering{\footnotesize{(b) PIRA's performance on pan-CDNs' traffic change.}}
    \end{minipage}
    \begin{minipage}{.32\linewidth}
        \centering
        \includegraphics[width=\textwidth]{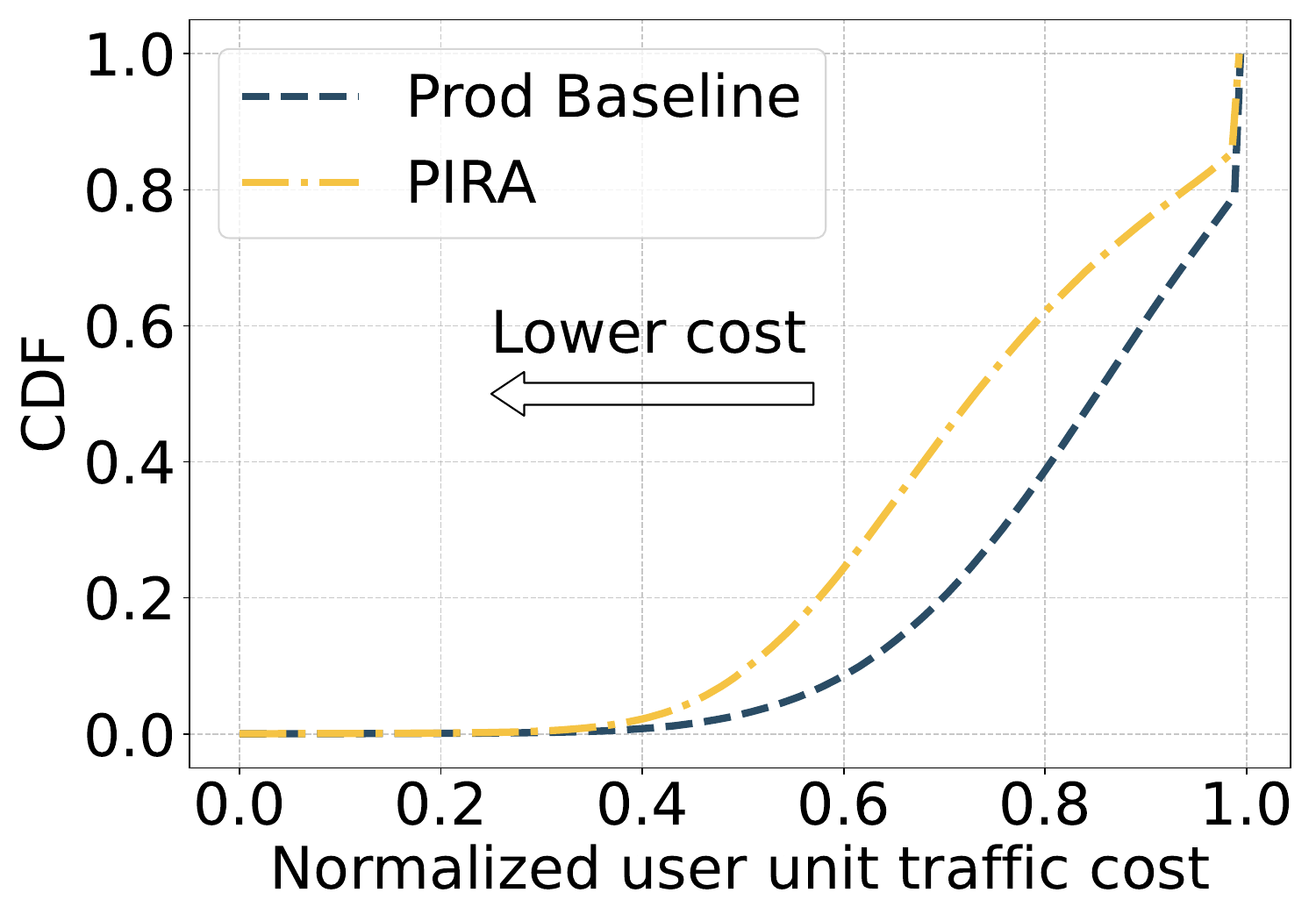}
        \centering{\footnotesize{(c) PIRA's performance on user-level unit traffic cost.}}
    \end{minipage}
    \vspace{-2pt}
    \caption{Results of the A/B experiment comparing PIRA against the production baseline strategy on the video platform.}
    \label{fig:online_qos}

\end{figure*}
\subsubsection{QoE performance.}
We calculate the rebuffering ratio, start-up delay and traffic cost in the simulations.
The experimental results are shown in Figure~\ref{exp:v1} and~\ref{exp:sd}.
Pure pan-CDN4 strategy achieves the optimal traffic cost with a significant service quality sacrifice.
While pure pan-CDN1 strategy performs a good quality on QoE metrics, it does not perform better than PIRA.
PIRA outperforms the production baseline in both QoE metrics and traffic cost over three time periods.
Specifically, PIRA achieves a 32\% reduction in average rebuffering ratio and a 24\% decrease in cost.
Moreover, PIRA achieves nearly the same cost performance as the theoretical optimal method Oracle, with the only gap arising from rebuffering caused by network throughput uncertainty.

\subsubsection{Efficiency analysis.}
We compare the performance and computation overheads of PIRA with/out the pruning techniques.
As shown in Figure~\ref{exp:pruning}, removing the pruning strategies can lead to up to a 2750× increase in computational overhead, while yielding less than a 2.5\% improvement in performance.
We also tested the runtime speed of PIRA on an iOS device with a CPU clocked at 3.2 GHz, with an average step execution time of under 0.025 seconds.

\subsection{Production deployment}

\subsubsection{Deployment details.}
We implement PIRA on our video platform and conduct online experiments to evaluate its performance, incorporating domain-specific engineering techniques to enhance real-world deployment.
For example, network connections are maintained in a connection pool for future reuse and only closed after several minutes of inactivity, rather than immediately after each video segment download.
The online experiment involves 450,000 users, conducted at scale to validate PIRA's real-world performance.
Users are randomly assigned to control and experimental groups.
The control group uses the production baseline strategy, while the experimental group fully implements the PIRA framework, enabling real-time optimization of pan-CDN selection and download range allocation.

\subsubsection{Experiment results.}
The experiment is conducted over two weeks, with results shown in Figure~\ref{fig:online_qos}.
QoE metrics and traffic changes are analyzed at different times of the day.

\noindent\textbf{PIRA leads to improved playback QoE.} PIRA reduces the average video start-up delay (AvgStDelay) by $2.1\%$, while also achieving a $13.4\%$ reduction in average rebuffering count (AvgRebufCnt)—i.e., the number of playback stalls—and a $15.2\%$ reduction in average rebuffering ratio (AvgRebufRatio).
While switching pan-CDNs introduces transient throughput degradation due to connection re-establishment overheads, PIRA mitigates this effect using connection pool techniques, reducing the average bitrate (AvgBitrate) deviation to just $0.4\%$ compared to the baseline.
Although this minor bitrate adjustment may slightly impact perceived quality, the overall QoE is significantly improved by reductions in rebuffering and startup delays, as supported by prior work on streaming performance tradeoffs~\cite{dobrian2011understanding}.

\noindent\textbf{PIRA minimizes the consumption of premium pan-CDN resources.} To assess PIRA’s impact on traffic costs, we analyze traffic usage for individual pan-CDNs.
As shown in Figure~\ref{fig:online_qos} (b), PIRA reduces pan-CDN1's traffic consumption by over $20\%$ across all time periods, while increasing traffic usage of other pan-CDNs by $20-40\%$.
This redistribution reflects PIRA’s strategy of prioritizing cost-efficient CDNs for non-critical downloads and offloading high-cost CDNs (e.g., pan-CDN1) to essential segments.
We also measure the normalized unit traffic cost (scaled to $[0,1]$ for privacy), shown in Figure~\ref{fig:online_qos} (c).
PIRA reduces the average normalized unit traffic cost by $10\%$, from 0.833 to 0.746.
This improvement arises from PIRA’s ability to balance QoE needs with pan-CDN pricing tiers, ensuring premium resources are used only when necessary.

\section{Related work}\label{related_work}
\textbf{Short video streaming.} The rapid growth of short-video platforms like Douyin~\cite{personalized}, and Instagram Reels presents unique technical challenges for service providers.
To ensure optimal user quality of experience (QoE), prior research has focused on network bandwidth prediction \cite{sun2016cs2p, cellular,qiao2022tracedriven}, congestion control~\cite{fan2025echocc, zhang2024bbq}, multipath transport~\cite{hu2025cross}, ABR streaming \cite{yin2015control, situ, qiao2020beyond,zhang2023quty, chen2024soda}, and neural video codecs \cite{swift, li2023neural}.
At the same time, the surge in video traffic has driven up bandwidth costs, prompting research into preloading optimization \cite{zhang2020apl, gamora, Digitaltwin, qian2022dam, chao2022pdas,liu2025deload} and heterogeneous CDN selection \cite{jiang2016cfa, jiang2017pytheas, zhang2024venus, meng2022dig} to reduce transmission expenses. This work addresses the critical gap in real-time pan-CDN source selection during video playback, aiming to optimize the QoE-cost tradeoff through dynamic resource allocation.

\noindent\textbf{Content delivery networks.} Prior work has focused on optimizing CDN selection for video delivery. CFA~\cite{jiang2016cfa} predicts CDN quality via trace-driven analysis and select the appropriate one, while Pytheas~\cite{jiang2017pytheas} enhances selection through real-time exploration–exploitation. DIG~\cite{meng2022dig} uses inter-session features to predict rebuffering. These methods adopt a server-side predict-and-select approach.
In contrast, some providers now use heterogeneous edge resources (e.g., smart home devices, idle servers)~\cite{ye2024kepc,zhang2024enhancing,ma2024sdpush}, offering them at low cost but with increased network variability and degraded service quality.
This motivates the need for a new intra-session, dynamic CDN selection strategy that can adapt to fine-grained network dynamics.

\section{Conclusion}\label{conclusion}
This paper proposes PIRA, a network resource adaptation framework that optimizes QoE and traffic cost by leveraging performance and pricing diversity across pan-CDNs.
Extensive simulations and large-scale online experiments confirm PIRA's effectiveness, showing notable gains over the production baseline.
Future work includes exploring Deep-RL to enhance adaptability and integrating components like ABR and preloading to further balance quality and efficiency.

\begin{acks}
We sincerely appreciate the anonymous reviewers for their valuable feedback and suggestions.
We also thank our colleagues in the network engineering team—Xiaoyi Xu, Chen Hu, Hao Zhong, Zhengqing Gao, and others—for their significant support in the industrial deployment of PIRA on our video platform, Douyin.
This work is supported in part by the National Natural Science Foundation of China under No. 62472041 and 62225204, in part by the A3 Foresight Program of NSFC under Grant 62061146002.
\end{acks}

\label{pira:last-body-page}
\clearpage
\label{pira:first-reference-page}

\bibliographystyle{ACM-Reference-Format}
\balance
\bibliography{refpaper}

\end{document}